\documentclass[aps,prd,showpacs,floatfix,preprintnumbers,nofootinbib,
amssymb,amsmath]{revtex4}
\usepackage{subfigure}
\usepackage{graphicx,bm,color}
\usepackage{amssymb}

\begin{document}
\title{Shear Viscosity and Phase Diagram from
Polyakov$-$Nambu$-$Jona-Lasinio model}
\author{Sanjay K. Ghosh}
\email{sanjay@jcbose.ac.in}
\author{Sibaji Raha}
\email{sibaji@jcbose.ac.in}
\author{Rajarshi Ray}
\email{rajarshi@jcbose.ac.in}
\author{Kinkar Saha}
\email{saha.k.09@gmail.com}
\author{Sudipa Upadhaya}
\email{sudipa.09@gmail.com}
\affiliation{Center for Astroparticle Physics \&
Space Science, Block-EN, Sector-V, Salt Lake, Kolkata-700091, INDIA 
 \\ \& \\ 
Department of Physics, Bose Institute, \\
93/1, A. P. C Road, Kolkata - 700009, INDIA}


\begin{abstract}
We discuss a detailed study of the variation of shear viscosity,
$\eta$, with temperature and baryon chemical potential within the
framework of Polyakov$-$Nambu$-$Jona-Lasinio model. $\eta$ is found
to depend strongly on the spectral width of the quasi-particles present
in the model. The variation of $\eta$ across the phase diagram has
distinctive features for different kinds of transitions. These variations
have been used to study the possible location of the Critical End Point (CEP),
and cross-checked with similar studies of variation of specific heat.
Finally using a parameterization of freeze-out surface in heavy-ion
collision experiments, the variation of shear viscosity to entropy
ratio has also been discussed as a function of the center of mass
energy of collisions.

\end{abstract}
\maketitle

\section{INTRODUCTION}

The relativistic heavy ion collision experiments provide us with the
unique opportunity to understand the physics of strongly interacting
matter expected to be present in the universe after a few microseconds
of the big bang, and possibly present in the interior of neutron stars.
In the experiments, two heavy ions colliding at relativistic energies
are expected to form a fireball consisting of deconfined quarks and
gluons, popularly known as the Quark-gluon plasma (QGP). The search for
QGP is continuing for almost last thirty years using several generations
of higher energy accelerators such as BEVALAC, AGS, SPS, RHIC and LHC
while covering a large energy range of few AGeV to few ATeV. Various
observables, such as, $J/\Psi$  suppression \cite{Matsui} and strangeness
enhancement \cite{Koch_muller} had been proposed as signatures of such
state of matter. All such proposed signatures are based on medium's
properties which differed substantially in hadronic and quark phases.
Charmonium suppression, for example, was based on the properties
of deconfinement and plasma screening \cite{Matsui}, whereas,
strangeness enhancement was based on the chiral symmetry restoration,
which may be fully realized in QGP but only partially in a hadron gas
\cite{Koch_muller,Meyer}.

If the main interest lies in the identification of a new form of bulk
matter then it is essential to choose observables corresponding to
unique collective properties of this matter. For example radial, azimuthal
and longitudinal flow are some of the relevant observables in heavy ion 
collisions \cite{Stocker1}. In general such observables may be obtained 
from azimuthal Fourier components \cite{Ollitrault}, $v_n(y, p_T , N_p , h)$
 of triple differential inclusive distribution of hadrons which are 
selected based on their impact parameter range. 

The observation of elliptic flow in non-central heavy-ion collisions at
RHIC, may be considered as the most important evidence for
hydrodynamical behavior of QGP.  Elliptic flow occurs when the plasma
collectively responds to pressure gradients in the initial state.
Hydrodynamic evolution converts the initial pressure gradients to
velocity gradients in the final state. In a heavy-ion collision one
cannot control the deformation of the initial state. Instead, the
deformation of the plasma is determined by the shape of the overlapping
region of the colliding nuclei. This shape is governed by the impact
parameter $b$. The impact parameter can be measured on an
event-by-event basis using the azimuthal dependence of the spectra of
produced particles. Once the impact parameter direction is known, the
particle distribution can be expanded in Fourier components of the
azimuthal angle $\phi$. The Fourier coefficients ($v_2$, $v_4$ etc.)
carry information about the deformation of the final state. For example,
a positive $v_2$ implies the preferential emission of particles in the
short direction, i.e. the presence of elliptic flow. Since shear
viscosity $\eta$ is expected to oppose the elliptic flow (and reduce
$v_2$), it is necessary to incorporate $\eta$ in the analysis. Moreover
a dimensionless quantity $\eta/s$, where $s$ is the entropy density,
tells us about the actual behavior of the fluid. The ratio $\eta/s$
is akin to the inverse of Reynolds number.

 In general $\eta$ for a system of quasi-particles is expected to vary as
$\sim~\epsilon t_{mft}$, where $\epsilon$ is the energy density and
$t_{mft}$ is the average mean free time. The entropy density
$s \sim ~k_B n$, where $k_B$ is the Boltzmann constant and $n$ the
number density. Since $\frac{\epsilon}{n}$ is the average energy per
particle, considering uncertainty principle one would get a lower
bound on the product of $\epsilon/n$ and $t_{mft}$. In other words one
would get $\eta/s \ge \frac{\hbar}{k_B}$. For the strong
coupling limit of superconformal-QCD, Policastro {\it et al.}
\cite{policastro} found $\eta/s = \frac{\hbar}{4\pi k_B}$. On the other
 hand, Kovtun {\it et al.} \cite{Kovtun} conjectured
$\frac{\eta}{s}\geq \frac{\hbar}{4\pi k_B}$ to be the lower bound 
(KSS bound) for a wide class of systems. Interestingly such a finite but
low value of $\eta/s$ is found to be consistent with the analysis of
RHIC data through hydrodynamical simulations \cite{Huovinen,Niemi}.

Thus heavy ion experiments suggest formation of QGP with a behavior of
near perfect fluidity i.e. very low viscosity. There are different
(2+1)d \cite{SH,Romatschke,Luzum,Dusling} and (3+1)d
\cite{Schenke,Bozek} viscous hydrodynamic codes which are used to
estimate QGP viscosity from the experimental data using the elliptic
flow coefficient, $v_2$. The initial spatial deformation of the fireball
created in relativistic heavy-ion experiments is converted into final
state momentum anisotropies through hydrodynamic simulations. Viscosity
comes into play via the degradation of this conversion efficiency. As in
experimental detections only the final state hadrons are tracked, the
most efficient observable to be related to these studies is the charged
hadron elliptic flow, $v_2^{ch}$. Therefore the best description is
provided with the amalgamation of the viscous hydrodynamic approach to
the QGP phase and a microscopic description for re-scattering of late
hadronic stage. Such hybrid approaches include VISHNU \cite{Bass} which
incorporates VISH2+1 \cite{CShen,SH} algorithm with UrQMD cascade model
\cite{SABass} and McGill code which connects (3+1)d viscous
hydrodynamics to UrQMD. The pioneering study in this regard was carried
out by Luzum and Romatschke \cite{Luzum} using (2+1)d viscous
hydrodynamics. One of the points of uncertainty in these studies is
initial condition. Different initial conditions like in MC-KLN or
MC-Glauber lead to uncertainties in the values of
$(\frac{\eta}{s})_{QGP}$ by factor of 2 to 2.5 \cite{Heinz}. A recent
study carried out using MUSIC + UrQMD \cite{Ryu}, with the IP-Glasma
initial conditions \cite{Prithwish}, shows excellent match to
multiplicity and flow distributions at RHIC and LHC. In fluid dynamical
descriptions of the created fireball, shear viscosity to entropy density
ratio $\frac{\eta}{s}$ is usually taken to be temperature independent.
Predictions made in order to explain the azimuthal anisotropies of the
spectra, like the elliptic flow coefficient $v_2$ reveal very small
value for this $\frac{\eta}{s}~\sim$ 0.1
\cite{Romatschke,Luzum,Hirano,Song}. However, there are some works
including the temperature dependence of $\frac{\eta}{s}$ \cite{Denicol}
as well, where the authors have taken different parameterizations to get
a thorough understanding of the effect on elliptic flow as well as
higher harmonics. Lattice studies of transport coefficients of a gluon
plasma have been carried out using $16^3\times 8$ and $24^3\times 8$
lattices \cite{Nakamura} indicating an ideal fluid behavior of QGP.

There are different techniques which can be used for the  evaluation of
$\eta$ in strongly interacting systems, namely, the Relaxation Time
Approach \cite{Chakraborty}, the Chapman-Enskog method \cite{Polak} and
the Green-Kubo formalism \cite{Kubo}. In Relaxation Time Approach (RTA), it is
assumed that the collisional effects drive the perturbed distribution
function close to the equilibrium one with a relaxation time of the
order of the time required for particle collisions. On the other
hand, the Chapman-Enskog approximation is based on the fact that on
slight shift of the distribution function from its equilibrium value, 
the former can be expressed in terms of hydrodynamical variables and
their gradients. One advantage of Chapman-Enskog method is that one can
do a successive approximation to get results closer to Kubo formalism.
The Green-Kubo formalism relates linear transport coefficients to
near-equilibrium correlations of dissipative fluxes and treats them as
perturbations to local thermal equilibrium. A comparative study of the
three different methods has been carried out by Wiranata $et~al.$ \cite{Wiranata}.
In the varied cases considered in this paper, the Green-Kubo
technique is found to be more reliable. In \cite{Plumari} it has been
argued that while in the case of CE, variational method may yield
solutions with arbitrary accuracy depending on the order of
approximation, RTA has no control over its accuracy. The RTA result was
found to differ from that obtained using Green-Kubo formalism by factor
of 2. On the other hand, the CE method, already at 1st order, was found
to display satisfactory agreement with the Green-Kubo results.
Comparisons in the context of the non-relativistic hard sphere can be
found in classical literature \cite{McLennan}, from which it is inferred
that higher order approximations of the CE method approaches the
Green-Kubo one. Notwithstanding this fact the RTA method has often been
used  due to its simplicity. It has been used to evaluate $\eta$ for
two flavor matter in NJL model \cite{Sabya}. On the other hand, in
\cite{Weise}, a combination of large $N_c$ expansion and Kubo formalism
was used to calculate $\eta$ in two flavor NJL model. In the present
work we have used the framework of 2+1-flavored
Polyakov-Nambu-Jona-Lasinio (PNJL) model to study the behavior of
$\eta$ at finite temperature and density. The variation of $\eta$ with
temperature and density is then used to discuss the location of 
critical end point (CEP).

The present article is organized as follows. The outline of the
formalism adopted for this work is given in Section II followed by a
brief introduction to the PNJL model in Section III. In Section IV-A,
we look for the variational nature of $\eta$ with $T$ for different
choices of the spectral width $\Gamma$. Further, we compute $\eta$ as
a function of quark chemical potential $\mu_q$ for three different
choices of $T$, viz. one in the expected range of 1st order 
phase transition, other in cross-over range and finally the last one
beyond the cross-over region. We also discuss the variation of 
$\frac{\eta}{s}$ as a function of $\mu_q$ for a wide range of $T$.
In Section IV-B, variation of $\frac{\eta}{s}$ with $T$ at various
$\mu_q$ has been used to draw the phase diagram and identify the CEP
region. The location of CEP is further validated with the behavior of
the specific heat, $C_V$ in Section IV-C . In Section IV-D, we calculate
$\frac{\eta}{s}$ under different experimental conditions considering the
freeze-out parameterization. Finally, the results are summarized in
Section V.

\section{Kubo Formalism}

Kubo formalism, as mentioned earlier, calls for the spectral width of
the degrees of freedom of the system involved. This is realized through
the fact that, shear viscosity coefficient $\eta$ is related to
retarded correlators of energy-momentum (E-M) tensor i.e. 4-point functions
in Matsubara space. The energy-momentum function is defined as ::
$T_{\mu \nu}=i\bar{\psi} \gamma_{\mu} \partial_{\nu}
\psi~-~g_{\mu \nu}\mathcal{L}$. Making use of this energy-momentum
tensor, Kubo formula for shear viscosity reduces to the form
\cite{Weise},
\begin{equation}
 \eta(\omega)=\frac{1}{15T}\int_{0}^{\infty}dte^{i\omega t}
\int d\vec{r}(T_{\mu \nu}(\vec{r},t),T^{\mu \nu}(0,0)),
\label{eta1}
\end{equation}
where $T_{\mu \nu}$ is the $(\mu, \nu)$ component of the E-M tensor of
quark matter. The factor $15$ in the above equation comes from the
following identity \cite{Weise},
\begin{equation}
 \int d^3 x x_i^2 x_j^2 f(x^2)~=~\frac{1}{15}\int d^3x x^4 f(x^2).
\end{equation}

Starting from Eq.(\ref{eta1}) and neglecting the surface terms at
infinity, we arrive at the following expression of $\eta$ in terms of
the retarded correlators::
\begin{equation}
 \eta(\omega)~=~\frac{i}{\omega}[\Pi^R(\omega)~-~\Pi^R(0)].
\label{eta2}
\end{equation}
Following the steps of \cite{Weise,Iwasaki} we get,
\begin{equation}
 \eta~=~\frac{\pi}{15T}\int_{-\infty}^{\infty} d\varepsilon
\int \frac{d^3p}{(2\pi)^3}p_x^2 f_{\Phi}(1-f_{\Phi})
Tr[\gamma_2\rho(\epsilon,p)\gamma_2\rho(\epsilon,p)],
\label{eta3}
\end{equation}
where, $f_{\Phi}$ is the distribution function of the fermionic fields
in presence of the Polyakov loop $\Phi$ and $\rho(\omega,p)$ is the
spectral function defined as,
\begin{equation}
 \rho(\omega,p)~=~\frac{1}{2\pi i}(G^A(\omega,p)~-~G^R(\omega,p)),
\label{rho}
\end{equation}
with $G^{R/A}$ being the advanced and retarded Green's function
represented as,
\begin{equation}
 G^{R/A}~=~\frac{1}{\not{p}-M \pm isgn(p_0)\Gamma(p)}.
\label{greensfunction}
\end{equation}
In the conventional notation, $M$ is the quasi-particle mass and
$\Gamma(p)$, the spectral width. Incorporating simple algebraic
techniques and using Eq.(\ref{eta3}) and Eq.(\ref{rho}), we finally
arrive at,
\begin{equation}
 \eta[\Gamma(p)]=\frac{16N_cN_f}{15\pi^3T}\int_{-\infty}^{\infty}
d\varepsilon\int_{0}^{\infty}dpp^6\frac{M^2\Gamma^2(p)
f_{\Phi}(\varepsilon)(1-f_{\Phi}(\varepsilon))}
{(({\varepsilon}^2-p^2-M^2+\Gamma^2(p))^2+4M^2\Gamma^2(p))^2},
\label{eta4}
\end{equation}
where, $N_c$ and $N_f$ are the no. of colors  and flavors respectively.

\section{PNJL Model}

The Polyakov-Nambu-Jona-Lasinio (PNJL) Model
\cite{meisinger,fukushima,megias,ratti1,ray3} is a QCD-inspired
phenomenological model developed by coupling the Polyakov loop potential
to the Nambu-Jona-Lasinio (NJL) model. In the NJL model, the
interactions between quarks is accounted for by four quark terms which
respect the chiral symmetry of the Lagrangian. Spontaneous breaking of
chiral symmetry takes place due to dynamical generation of fermion mass. 
Since the gluons are integrated out, the NJL model fails to simulate the
deconfinement physics. Because of its failure to incorporate
confinement, some important aspects of the QCD thermal transition are
not suitably accounted for in the NJL model. However unlike NJL model,
the PNJL encapsulates this feature of the QCD transition. In this model,
the gluon dynamics is described by the background temporal field. So,
here the chiral and deconfinement order parameters are entwined into a
single framework. Several quantities like pressure, number density,
speed of sound etc. calculated in this model show very good agreement
with lattice results \cite{ray1,ray2,ray3}. In the present work we have
considered $SU(3)_f$ version of PNJL model including 8-quark
interaction \cite{deb1}. The current quark masses used here are
$m_u = m_d =$ 8.7 MeV and $m_s=$179.5 MeV along with the three momentum
cutoff $\Lambda=$640 MeV \cite{deb1}.

While computing the shear viscosity $\eta$ as in Eq.(\ref{eta4}), we
incorporated the modified Fermi-Dirac distribution functions
($f_{\Phi}$) in which the effect of the background Polyakov loop fields
is taken into account. The forms of the distribution functions for the
particles and anti-particles as realized in PNJL model are ::
\begin{equation}
 f_{\Phi}^{+}(E_p)=\frac{(\bar{\Phi}+
2\Phi e^{-\beta(E_p+\mu)})e^{-\beta(E_p+\mu)}+
e^{-3\beta(E_p+\mu)}}
{1+3(\bar{\Phi}+\Phi e^{-\beta(E_p+\mu)})e^{-\beta(E_p+\mu)}+
e^{-3\beta(E_p+\mu)}},
\end{equation}
\begin{equation}
 f_{\Phi}^{-}(E_p)=\frac{(\Phi+
2\bar{\Phi} e^{-\beta(E_p-\mu)})e^{-\beta(E_p-\mu)}+
e^{-3\beta(E_p-\mu)}}
{1+3(\Phi+\bar{\Phi} e^{-\beta(E_p-\mu)})e^{-\beta(E_p-\mu)}+
e^{-3\beta(E_p-\mu)}},
\end{equation}
where, '$\pm$' refer to the particle and anti-particle respectively.
$\Phi~\&~\bar{\Phi}$ are the Polyakov loop fields, $\mu$ and $\beta$
being the chemical potential and inverse temperature respectively.
Starting from PNJL thermodynamic potential we can calculate the fields
(degrees of freedom of the system involved), pressure and constituent
masses at corresponding temperatures and chemical potentials. The
details of the technique for 2 and 2+1 flavor cases can be found in
\cite{ratti1,ray3} and  \cite{ciminale,deb1} respectively. With these
inputs, we proceed to determine $\eta$ and $s$. 

\section{Results}
\subsection{Parameterization of $\Gamma$}

\begin{figure}[!h]
 {\includegraphics[height=6cm,width=8.0cm,angle=0]
{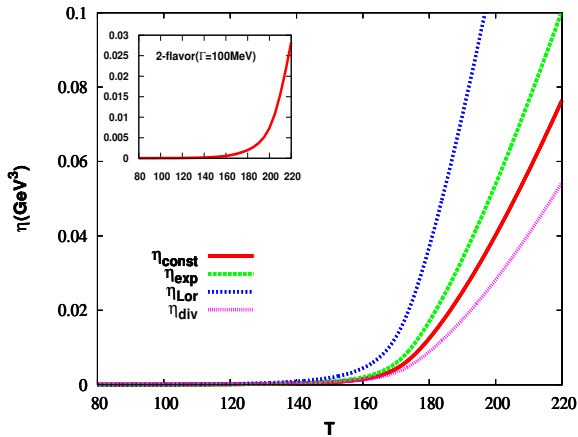}}
 \caption{(color online)$\eta$ as a function of temperature at vanishing
chemical potential for different forms of $\Gamma$}. 
\label{etaTdiffgamma}
\end{figure}

The spectral width $\Gamma$(p), in PNJL model, is supposed to have
contributions from Landau damping of quarks and mesons along with the 
recombination processes {\it i.e.} the formation of collective mesonic
modes due to quark - antiquark rescattering. Spectral width for 2 flavor
case including both $\sigma$ and $\pi$ has been evaluated at
next-to-leading order in the large $N_c$ expansion including one-loop
mesonic contributions \cite{Weise,Muller}. Spectral widths evaluated
this way would depend on both $T$ and $\mu$. In the case of $SU(3)_f$
PNJL model all the scalar and pseudoscalar meson channels will
contribute to this process. Moreover, the decay widths themselves may
become comparable or larger than mass, especially at lower temperatures.
Under such circumstances, one should really express $\eta$ explicitly in
terms of spectral functions which should then be evaluated considering
all possible channels \cite{Albericohtl}. Since such a calculation is
extremely involved, in our present work, we have considered the forms
of $\Gamma(p)$ as given in \cite{Weise}. The configurations of $\Gamma$
proposed to ensure the convergence of $\eta$ are::
\begin{center}
$Constant~~:\Gamma_{const}=100$MeV,\\
$Exponential:\Gamma_{exp}=\Gamma_{const}e^{-\beta p/8}$,\\
$Lorentzian:\Gamma_{Lor}(p)=\Gamma_{const}\frac{\beta p}
{1+(\beta p)^2}$,\\
$Divergent:\Gamma_{div}(p)=\Gamma_{const}\sqrt{\beta p}$.\\
\end{center}
Fig.(\ref{etaTdiffgamma}) shows the variation of $\eta$ with $T$, at
$\mu_q$=0, for different choices of $\Gamma$ as given earlier, whereas in
the inset result considering 2 flavors for a constant $\Gamma$=100 MeV
is shown. For a given $\Gamma$, $\eta$ is found to increase with
temperature. This behavior is similar to a gaseous system where
viscosity increases with temperature due to the increase in the average
momentum of the particles \cite{Schafer}. In the present case, the
increase in $\eta$ may be attributed to the decrease in the quark mass
with temperature in our model.  Fig.(\ref{etaTdiffgamma}) shows that
$\eta$ becomes very small in the low temperature range. In this low
temperature region, the quark masses become large and $\eta$ is expected
to fall as $M^{-6}$ \cite{Weise}. It can also be seen that
$\eta_{Lor}~>~\eta_{exp}~>~\eta_{const}~>~\eta_{div}$. This behavior
simply depends on the value of $\Gamma$ at a given temperature. A lower
value of $\Gamma$ corresponds to weaker interaction and hence a larger
mean free path \cite{Weise}. It is evident fom Eq.(\ref{eta4}) that the 
$\eta$ for 2-flavor matter will be less than the 3-flavor for equal masses. 
Since, $s$ quark mass is higher than the $u$ and $d$ masses, 
the difference is less than the equal mass case. In the inset of 
Fig.(\ref{etaTdiffgamma}) we have shown $\eta$ for 2-flavor matter 
evaluated in PNJL with the parameter set adopted in \cite{ray1,ray2}.  

\begin{figure}[!h]
 {\includegraphics[scale=0.28,angle=270]{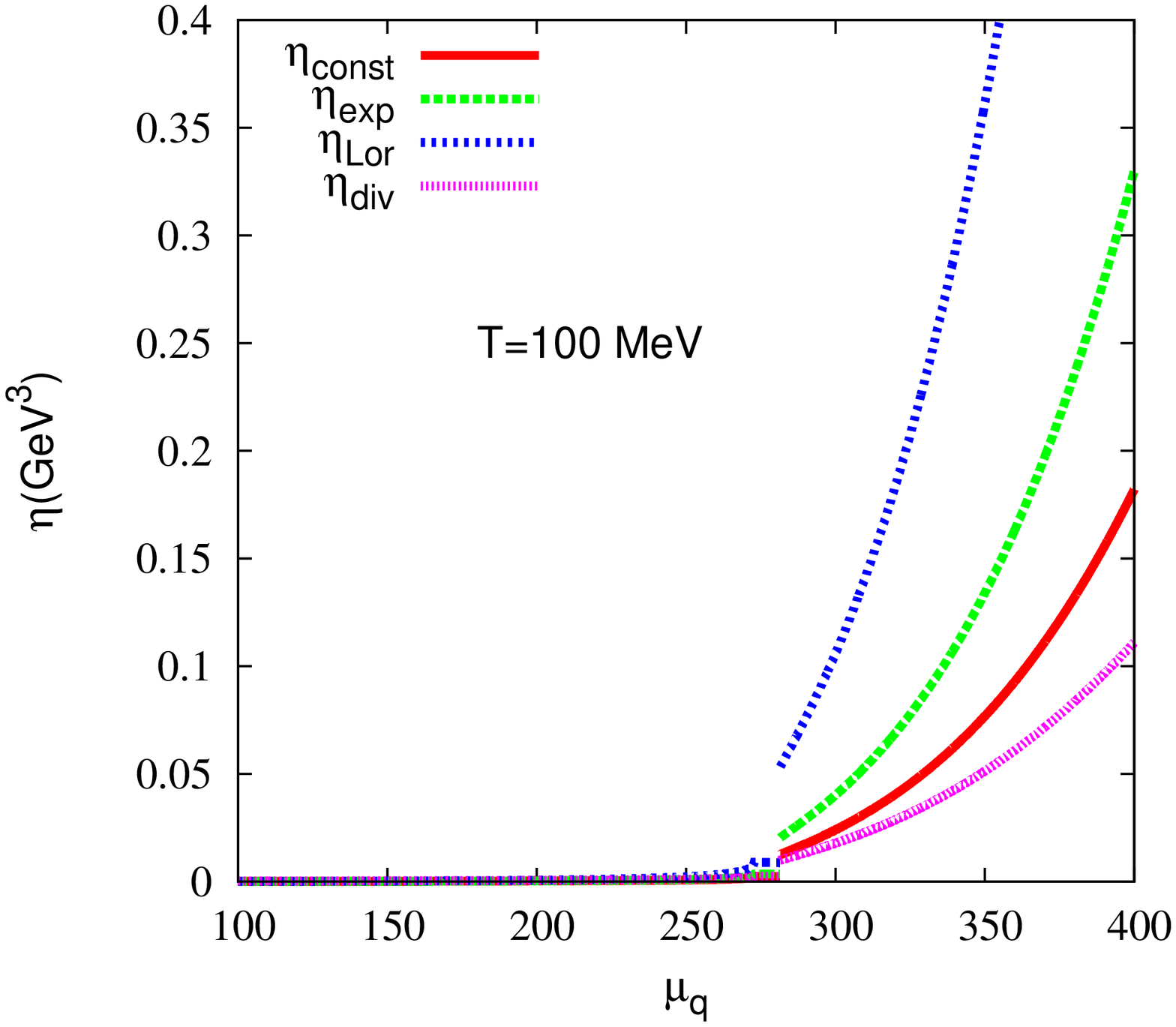}
 \includegraphics[scale=0.28,angle=270]{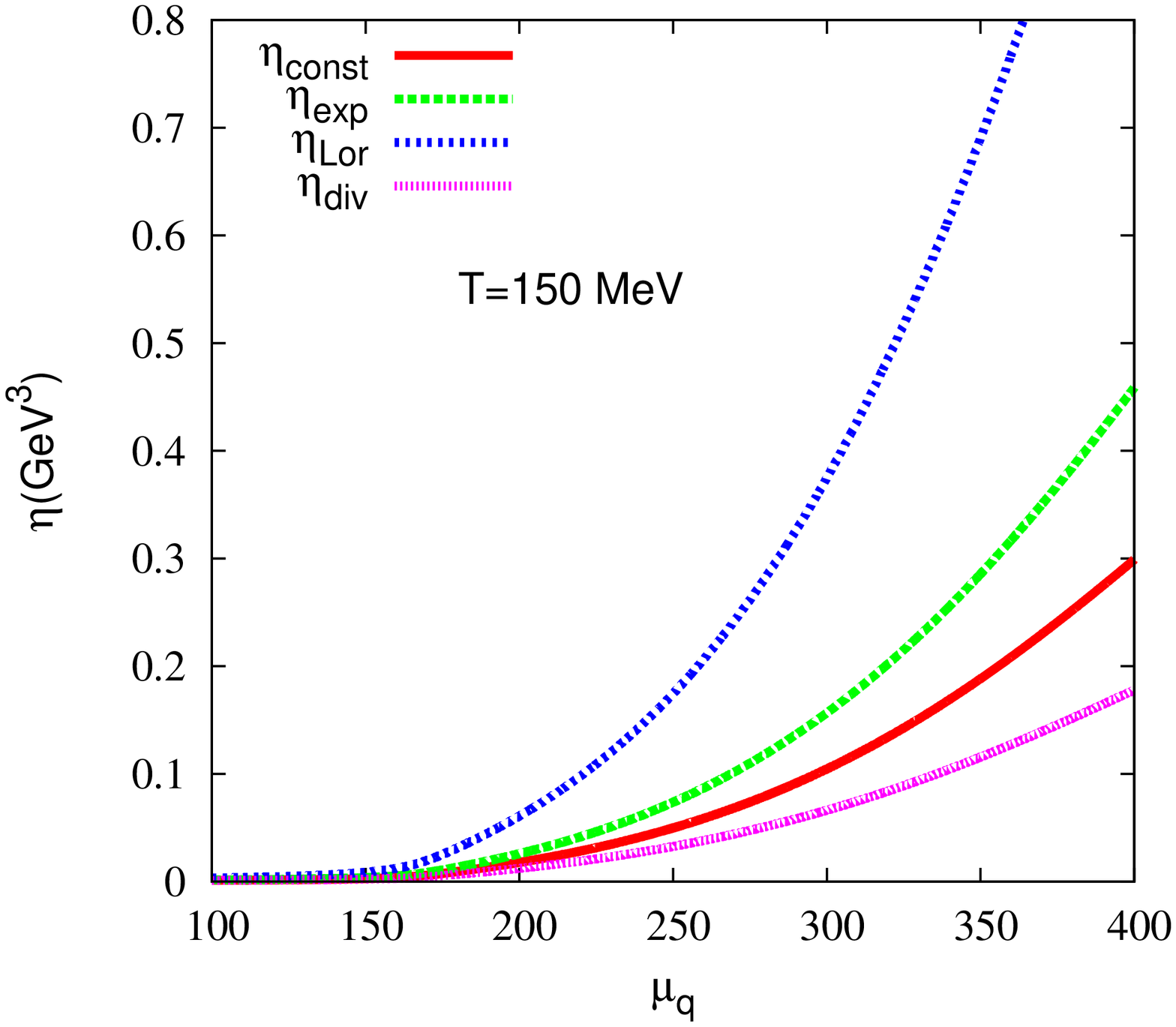}
 \includegraphics[scale=0.28,angle=270]{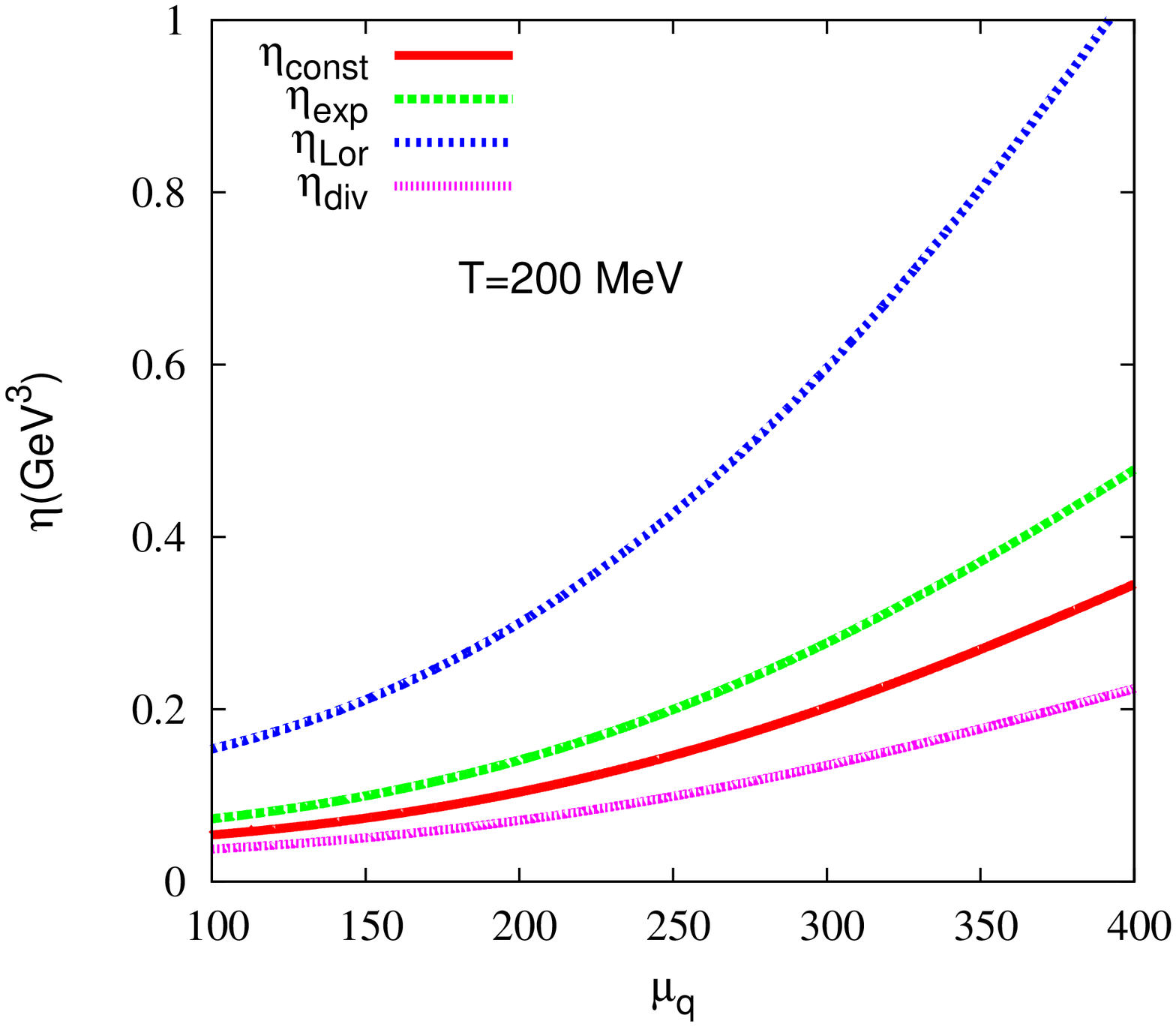}}
 \caption{(color online)$\eta$ as a function of quark chemical potential
at fixed temperatures for different forms of $\Gamma$}. 
\label{etamuqdiffgamma}
\end{figure}

Variation of $\eta$ with $\mu_q$ is shown in
Fig.(\ref{etamuqdiffgamma}). Here we have chosen three different
temperatures T= 100, 150 and 200 MeV, corresponding to the 1st order
phase transition, cross-over and beyond cross-over regions. Similar to
the zero chemical potential case, Fig.(\ref{etamuqdiffgamma})  also
shows increase in $\eta$ with $\mu_q$ at fixed $T$. However, the nature
of the curves are different for the three different choices of
temperature. For T=100 MeV, moving along the $\mu$-axis, one is expected
to encounter the 1st order phase transition line. Here, $\eta$ shows a
jump for all forms of $\Gamma$ at $\mu_q$ $\simeq$ 280 MeV. On the other
hand, both for T=150 MeV and T=200 MeV Fig.(\ref{etamuqdiffgamma})
shows, as expected, a smooth variation in $\eta$ along $\mu$-direction.
Moreover, $\eta$ at finite $\mu_q$ is larger than that for $\mu_q$=0 and
increases substantially for larger $T$. 

It has already been mentioned that smaller $\Gamma$ correspond to a
larger $\eta$. Hence the different forms of $\Gamma$, as observed
earlier, only affect the rate at which $\eta$ changes with $T$ and
$\mu_q$. But these forms do not change the qualitative behavior of
$\eta$, as seen from Figs.(\ref{etaTdiffgamma})-(\ref{etamuqdiffgamma}).
As mentioned earlier, in an explicit calculation one should evaluate
the spectral width $\Gamma$(p) considering the contributions from all
meson channels in $SU(3)_f$ PNJL model. This $\Gamma$(p) will depend on
both $T$ as well as $\mu$. In the absence of such a rigorous evaluation,
we would use the sum rule essential for the choice of Breit-Wigner form,
as a guiding principle \cite{Alberico,Iwasaki},
\begin{equation}
 \frac{1}{4}Tr_{spin}\int_{-\infty}^{\infty}\frac{d\epsilon}{2\pi}
[\rho(\epsilon,p)\gamma_0]~=~1.
\label{sumrule}
\end{equation}
The Eq.(\ref{sumrule}) is satisfied approximately when the value of
$\Gamma \le M$. So for our further discussions we choose a constant 
spectral width $\Gamma_{const}$=100 MeV so that the sum rule is
in general violated to a considerable extent only in the low momentum
region. In fact, as the quark masses drop sharply around $T_c$, from
its $T$ = $\mu_q$ = 0 value, the violation is expected to be more 
pronounced for $T~>~T_c$ only. But again the smaller contribution from
low momentum at these high temperatures keeps the violation to a minimum
level. For $\Gamma~>$ 100 MeV, $\eta$ variation is similar to that for
$\Gamma_{div}$, whereas, for $\Gamma~<$ 100 MeV, it is similar to that
for $\Gamma_{Lor}$. At the same time, in the regime of momentum transfer
comparable to QCD-scale($\sim$ 200 MeV), $\Gamma(p)$ becomes $\sim$ 100
MeV considering contributions from one-loop mesonic channels at
next-to-leading order in the large $N_c$ expansion as shown in
\cite{Weise}. There, the authors have included Landau damping and
recombination process as the leading dissipative effects to calculate
shear viscosity at the one loop level taking all mesons influencing the
spectral width. However going further up in the quark-momentum leads
towards decrease in spectral width which in turn increases the shear
viscosity.
\begin{figure}[!h]
 {\includegraphics[height=8cm,width=6cm,angle=270]{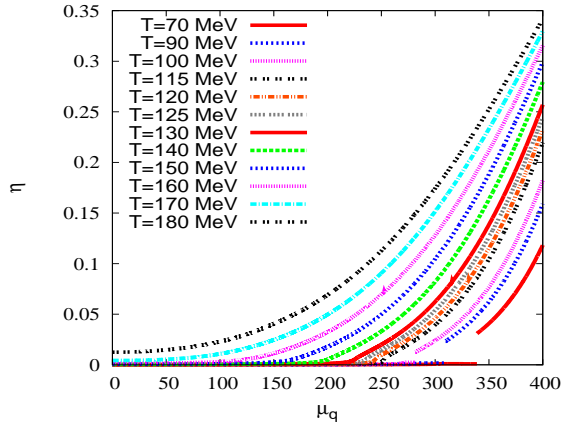}}
 \caption{(color online)Variation of shear viscosity with quark chemical
potential at different temperatures}. 
\label{etamuqdiffT}
\end{figure}

Let us again look at the variation of $\eta$ with $\mu_q$ for different
values of $T$ with $\Gamma$=100 MeV. From Fig.(\ref{etamuqdiffT}), two
distinct regimes are clearly visible. For T= 70 to 100 MeV, we get a
jump in $\eta$, whereas for T = 120 to 180 MeV, $\eta$ has a smooth
behavior analogous to the results displayed in Fig.(\ref{etamuqdiffgamma}). 
The temperature range T = 70 - 100 MeV and 120 - 180 MeV lie in the 
first order and cross-over zones of transition respectively and
Fig.(\ref{etamuqdiffT}) shows anticipated outcome of $\eta$ with
$\mu_q$. The region around T $\simeq$ 115 MeV calls for special
attention and will be discussed in details in the upcoming section.

\subsection{$\eta$/s $\&$ the Phase Diagram}

\begin{figure}[!h]
 {\includegraphics[height=7cm,width=9.5cm,angle=0]
{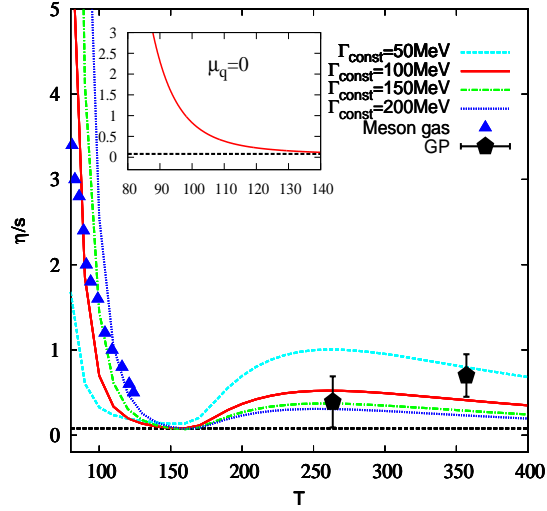}}
 \caption{(color online)Specific shear viscosity ($\frac{\eta}{s}$) as a
function of temperature at vanishing chemical potential}. 
\label{etabysT-mu0}
\end{figure}

In general $\eta$ for different fluids vary widely differing by orders
of magnitudes \cite{Schafer}. In such circumstances, Reynolds number,
the ratio of inertial to viscous forces in the Navier-Stokes equation,
is traditionally used as a measure of fluidity. In the case of
relativistic fluids, the Reynolds number (more specifically its inverse)
may be defined in terms of $\eta$/s, s being the entropy density. Here
we present this ratio in Fig.(\ref{etabysT-mu0}) for $\mu_q~=~0$ for 2+1
flavor quark matter in PNJL model. We have shown comparative plots for
different constant values of the spectral width to get a better
understanding of the effect of $\Gamma$ on the specific shear viscosity, 
$\frac{\eta}{s}$. The results are compatible with the arguments presented 
in the last section. $\frac{\eta}{s}$ has been evaluated in NJL model 
in \cite{Weise} using Kubo formalism whereas in \cite{Sabya} the ratio
was obtained using relaxation time approach. Our results in the PNJL
model are qualitatively similar to those obtained in \cite{Weise}.

As shown in Fig.(\ref{etabysT-mu0}) $\frac{\eta}{s}$ starts from high
values at low temperatures and decreases to a minimum
of $\frac{1}{4\pi}$ corresponding to ideal fluids near $T = T_c$. As
discussed earlier, $\eta$ itself is extremely small for $T<T_c$.
The behavior of $\frac{\eta}{s}$ for this region of temperature is
due to a larger drop in entropy density of the system. A simple
calculation for pion gas shows $\frac{\eta}{s}$ to be proportional to
$(\frac{f_{\pi}}{T})^4$ where, $f_{\pi}$ is the pion decay constant
\cite{Kapusta,Prakash}.  Hence, for T$\rightarrow$0, $\frac{\eta}{s}$
should diverge. Similar results have also been obtained by Lang
$et.~al.$ \cite{Kaiser} who have computed $\frac{\eta}{s}$ for
interacting pion-gas considering different pion masses. In the inset of
Fig.(\ref{etabysT-mu0}), the region from T=80 to 140 MeV has been zoomed
in for a better comparison with the system of interacting pion gas
\cite{Kaiser}. A comparison of our results in this temperature region
with corresponding result for meson gas ($T<T_c$) obtained from chiral
perturbation theory \cite{Chen} is also shown in
Fig.(\ref{etabysT-mu0}).

In general, for an ideal gas of quarks, the $\eta$ and hence
$\frac{\eta}{s}$ should diverge at large $T$ limit. As shown in
Fig.(\ref{etabysT-mu0}) $\frac{\eta}{s}$ does show an increasing
trend for $T>T_c$. The behavior of $\frac{\eta}{s}$ up to about
$T \sim 1.5 T_c$ seems to closely resemble the features of a fluid
having liquid-gas phase transition, for which a minima is expected
 near the transition point \cite {Schafer,Frenkel}.
However, since all the quark masses drop to their respective current
masses for $T>1.5T_c$, entropy starts dominating. As a result
$\frac{\eta}{s}$ starts decreasing slowly with increasing $T$, and
hence displays a behavior of an interacting liquid. In lattice studies
similar behavior has been observed for pure glue plasma (GP) as shown
in Fig.(\ref{etabysT-mu0}). The results of perturbation theory
\cite{Arnold,Blaizot} is around 1 as both $\eta$ and $s$ varies as
$T^3$ at higher temperatures. Eventually for asymptotic temperatures
an ideal gas behavior is expected to be restored.

\begin{figure}[!h]
 {\includegraphics[scale=0.30,angle=270]{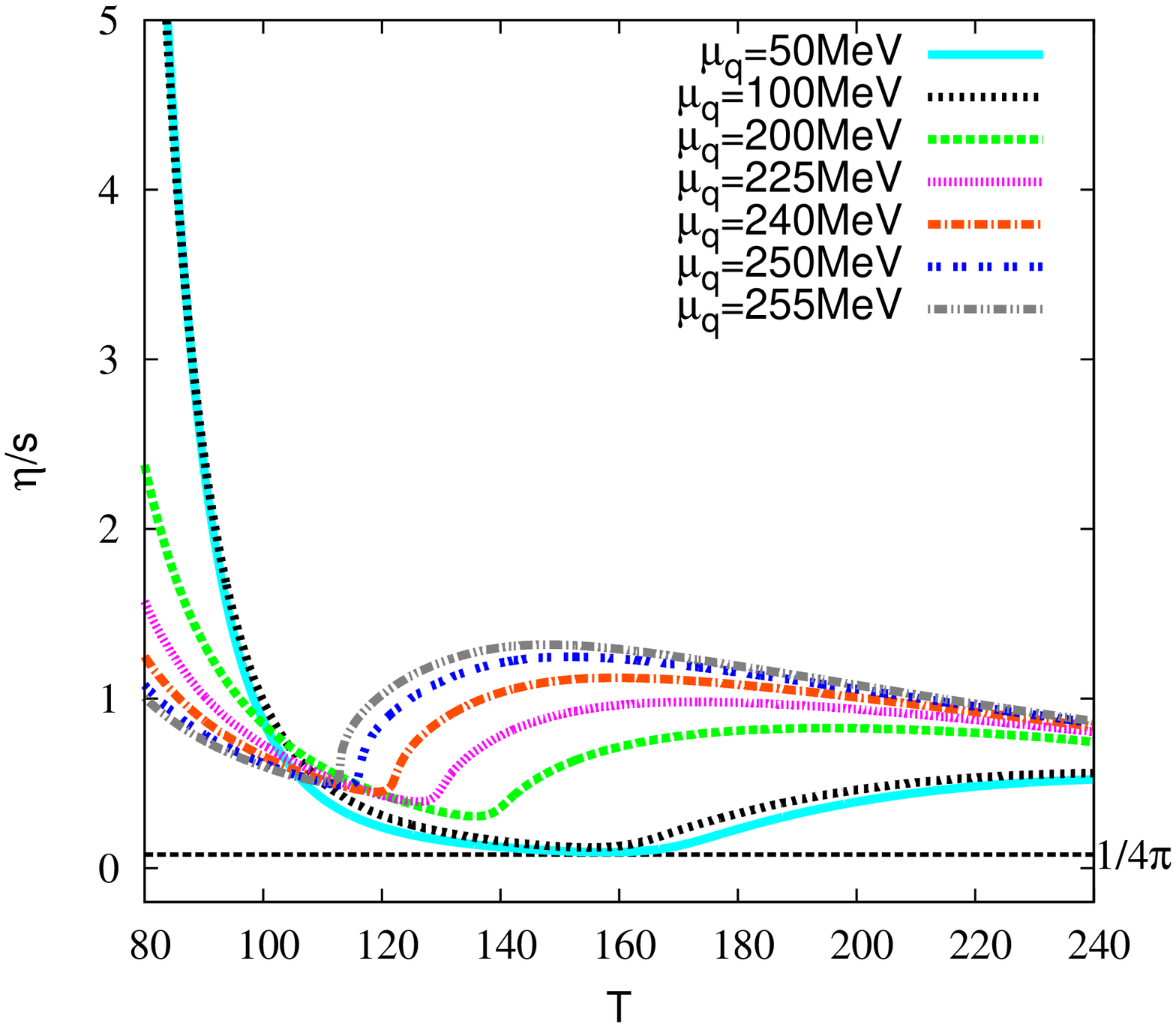}
 \includegraphics[scale=0.30,angle=270]{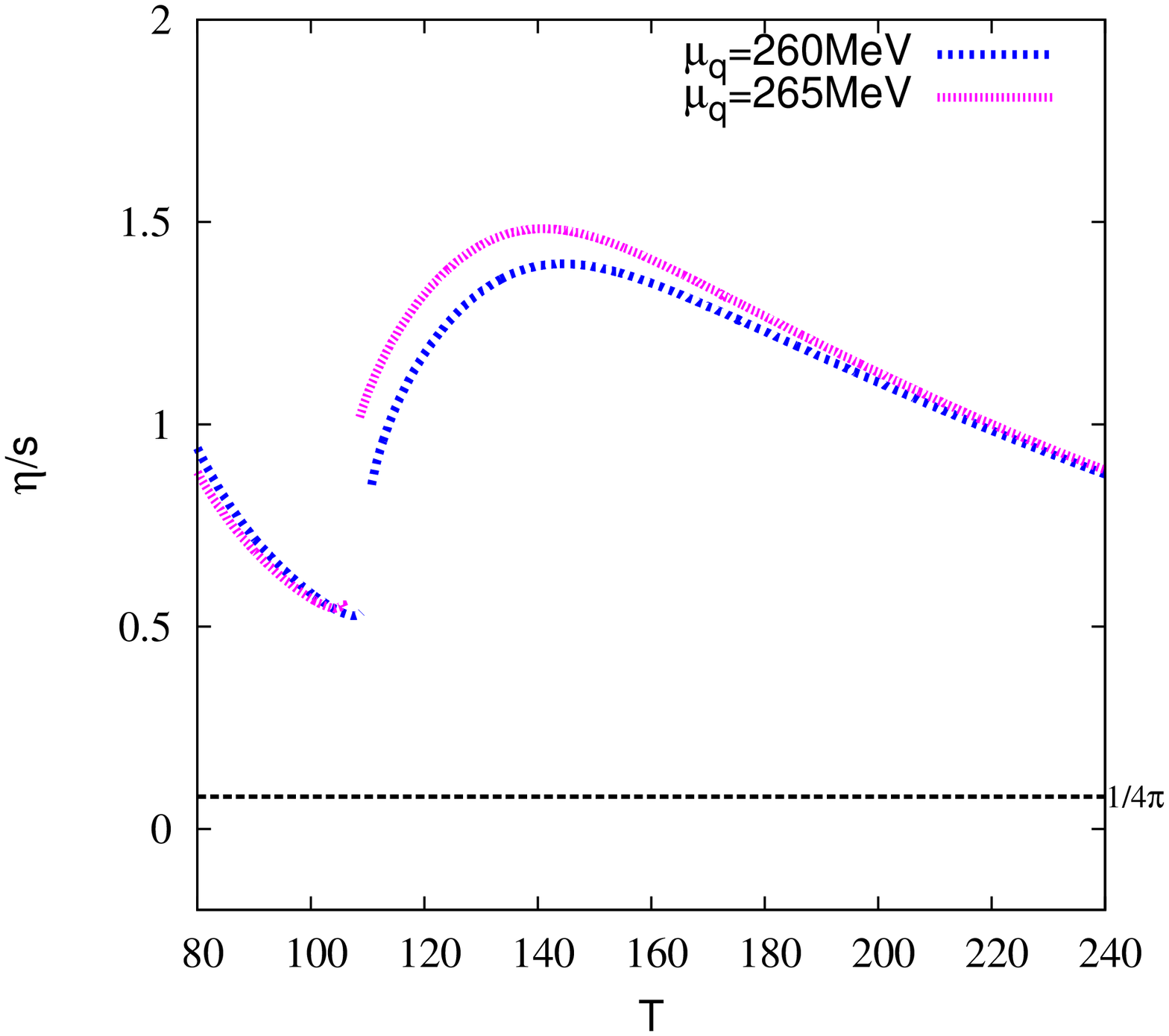}}
 \caption{(color online)$\frac{\eta}{s}$ as a function of temperature
for different non-zero values of quark chemical potential}. 
\label{etabysT}
\end{figure}

Similar features are observed with the increase in $\mu_q$ as shown in
Fig.(\ref{etabysT}). To summarize the situation we see that for
$\mu_q$=0, $\frac{\eta}{s}$ initially decreases with increase in $T$,
reaches the minimum of $\frac{1}{4\pi}$ bound near $T_c$. Thereafter it
increases with $T$ rather slowly after transition before stabilizing/
slowly decreasing in accordance with the behavior found in
Ref.\cite{Lacey,Nakamura,HBMeyer}. As we increase $\mu_q$, the same
feature is observed for $\mu_q \le$ 100-150 MeV. On the other hand,
for $\mu_q \ge$ 200 MeV, the system undergoes a transition at lower
temperatures and $\frac{\eta}{s}$ has a minimum which is significantly
higher than the KSS bound. Moreover, for $\mu_q \ge$ 260 MeV,
$\frac{\eta}{s}$ shows a jump which may be attributed to a first
order phase transition. Therefore a possibility of observing a
CEP arises near these ranges of $T$ and $\mu_q$.

QCD critical end point is supposed to be described by model H
\cite{Sonprd70}. ${\eta}$ is expected to diverge, with a
small power of correlation length, near the CEP
\cite{Hohenbergrmp49,Onukipre55}. On the other hand, for weakly coupled
real scalar field theories, $\frac{\eta}{s}$ is most likely to develop a
cusp at CEP \cite{Huangplb670}. A discontinuity in the behavioral
pattern of $\frac{\eta}{s}$ around CEP region has also been discussed in
\cite{Sasaki}.

In our study the location of the minima and the discontinuities 
of $\frac{\eta}{s}$ in Fig.(\ref{etabysT}), enable us to extract the
critical values of $T~\&~\mu$ to draw the Phase-diagram. However in the
very low temperature region extrapolation has been done considering the
fitting function in the form of polynomial 
\begin{equation}
 T = a_0 + a_1\mu + a_2 \mu^2
\end{equation}
with, $a_0~=~50~ MeV$, $a_1~=~-2.5$ and $a_2~=~-0.04~MeV^{-1}$. The
phase diagram along with the CEP region (black dot) has been plotted in
Fig.(\ref{etabysT}). 
\begin{figure}[!h]
 {\includegraphics[height=8.5cm,width=6cm,angle=270]
{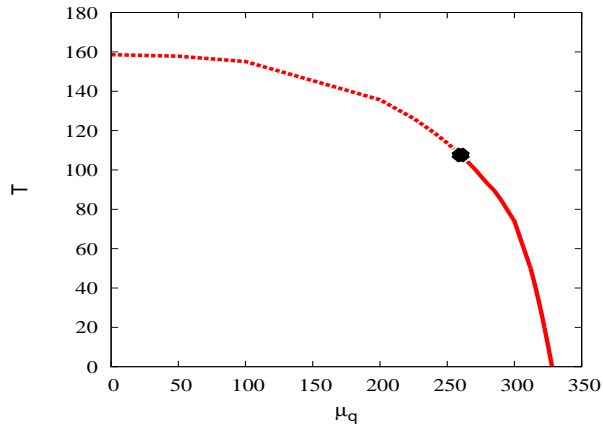}}
 \caption{(color online)Phase diagram for 2+1 flavor PNJL Model}. 
\label{phasediagram}
\end{figure}

The understanding of the behavior of the strongly interacting system
near CEP along with its location is extremely important.
Various efforts are being undertaken, both theoretically as
well as experimentally, to determine the position of CEP
\cite{Lacey,Ajitanand}. In general, second order derivatives of
thermodynamic quantities are expected to diverge near CEP which
is a second order transition point. These quantities may provide
additional information regarding the CEP. Here, in the PNJL model,
we have observed that at or around CEP, wide variations in order
parameters, fluctuations of conserved charges like net electric charge
\cite{SK} can occur as far as dynamic as well as static properties of
the system are concerned. 

\subsection{On the location of CEP}
In order to verify our results, we have considered the specific heat
$C_V$, which is expected to show diverging behavior near the CEP.
One can define $C_V$ as ::
\begin{equation}
 C_V = \frac{\partial\epsilon}{\partial T} = T \frac{\partial ^2 P}
{\partial T^2} = T \frac{\partial s}{\partial T}
\end{equation}
The divergence in $C_V$ near the CEP will translate into highly enhanced
transverse momentum fluctuations or highly suppressed temperature
fluctuations for a system passing close to the CEP.
\begin{figure}[!h]
 {\includegraphics[scale=0.28,angle=270]{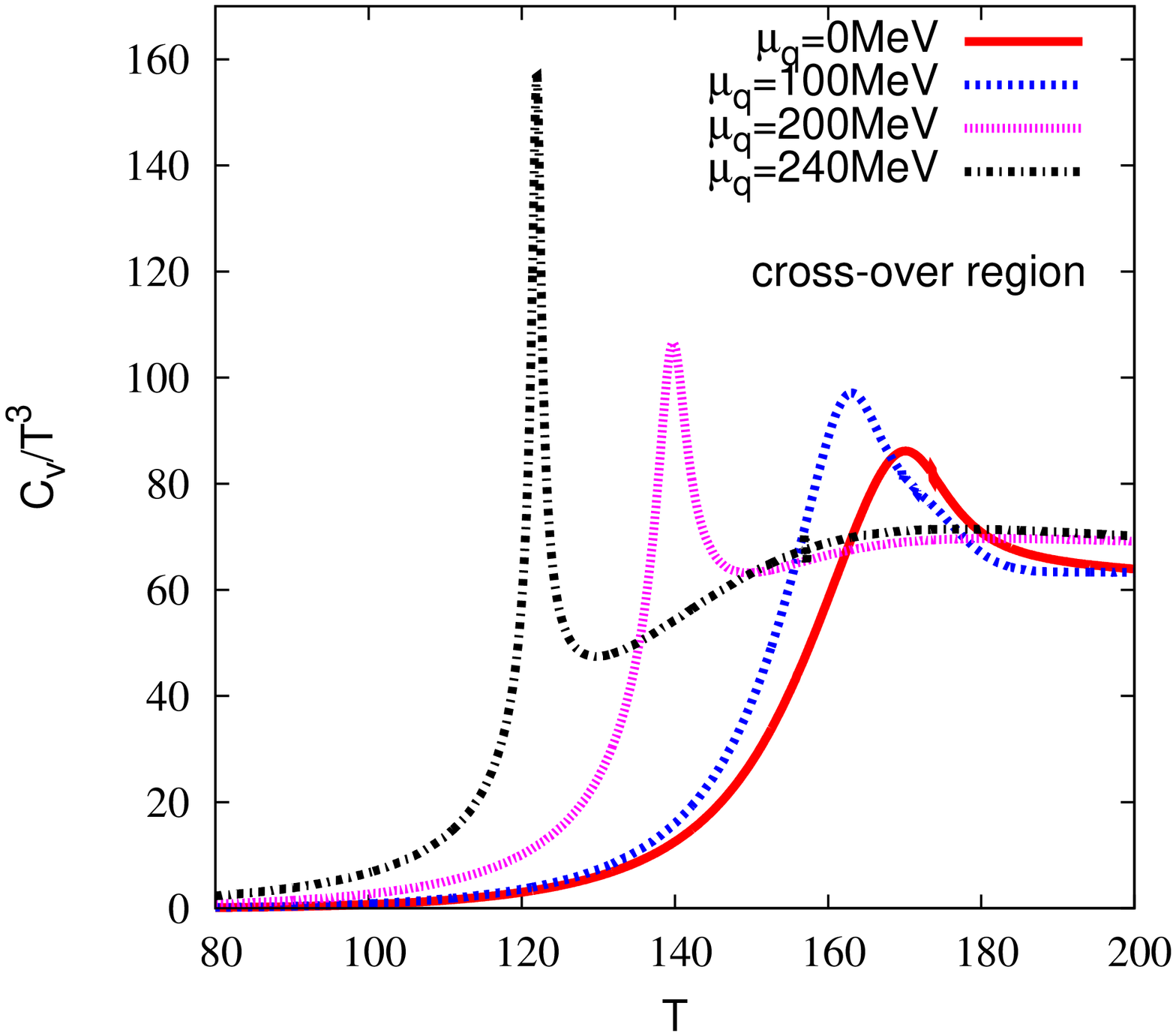}
 \includegraphics[scale=0.28,angle=270]{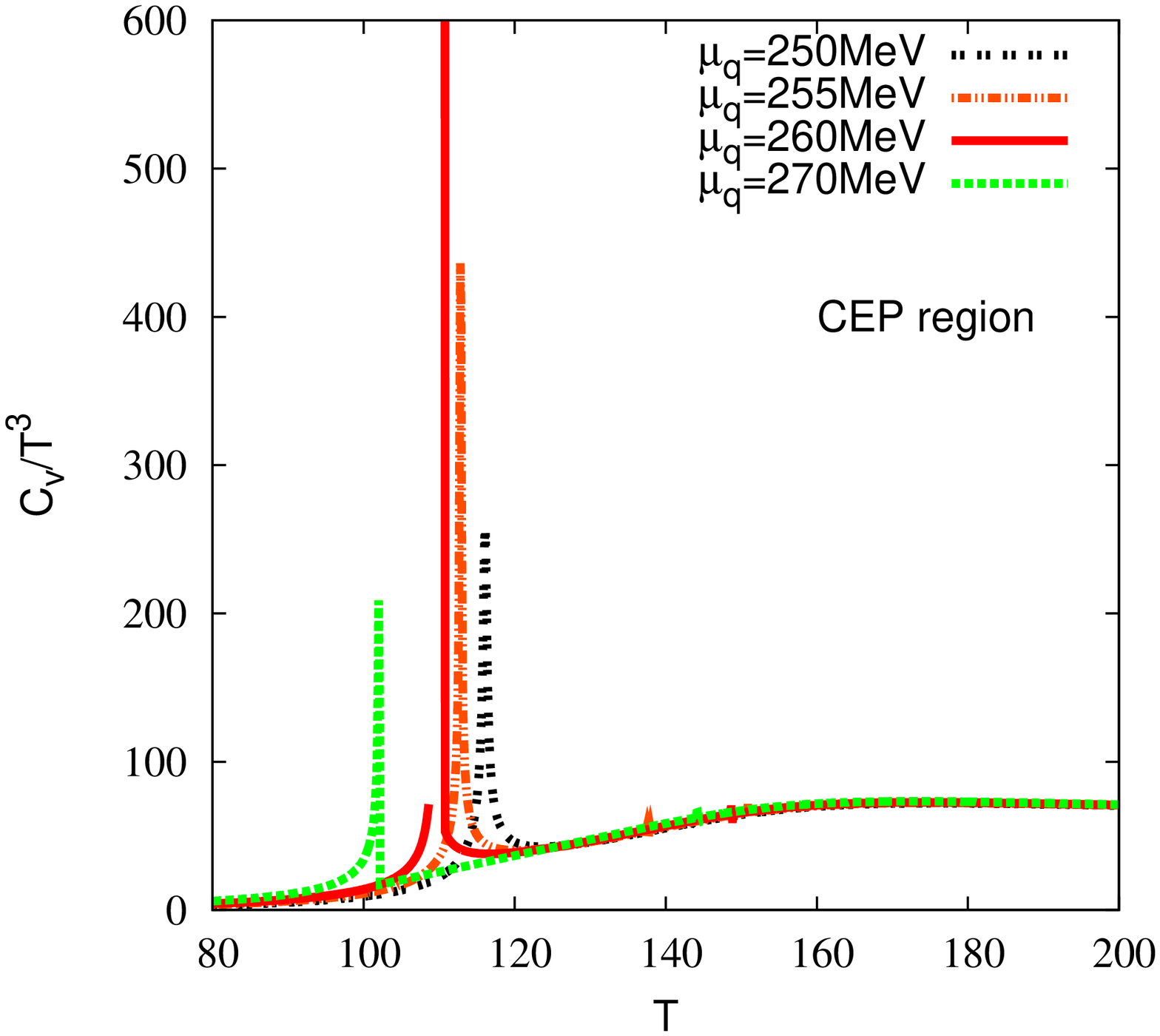}
 \includegraphics[scale=0.28,angle=270]{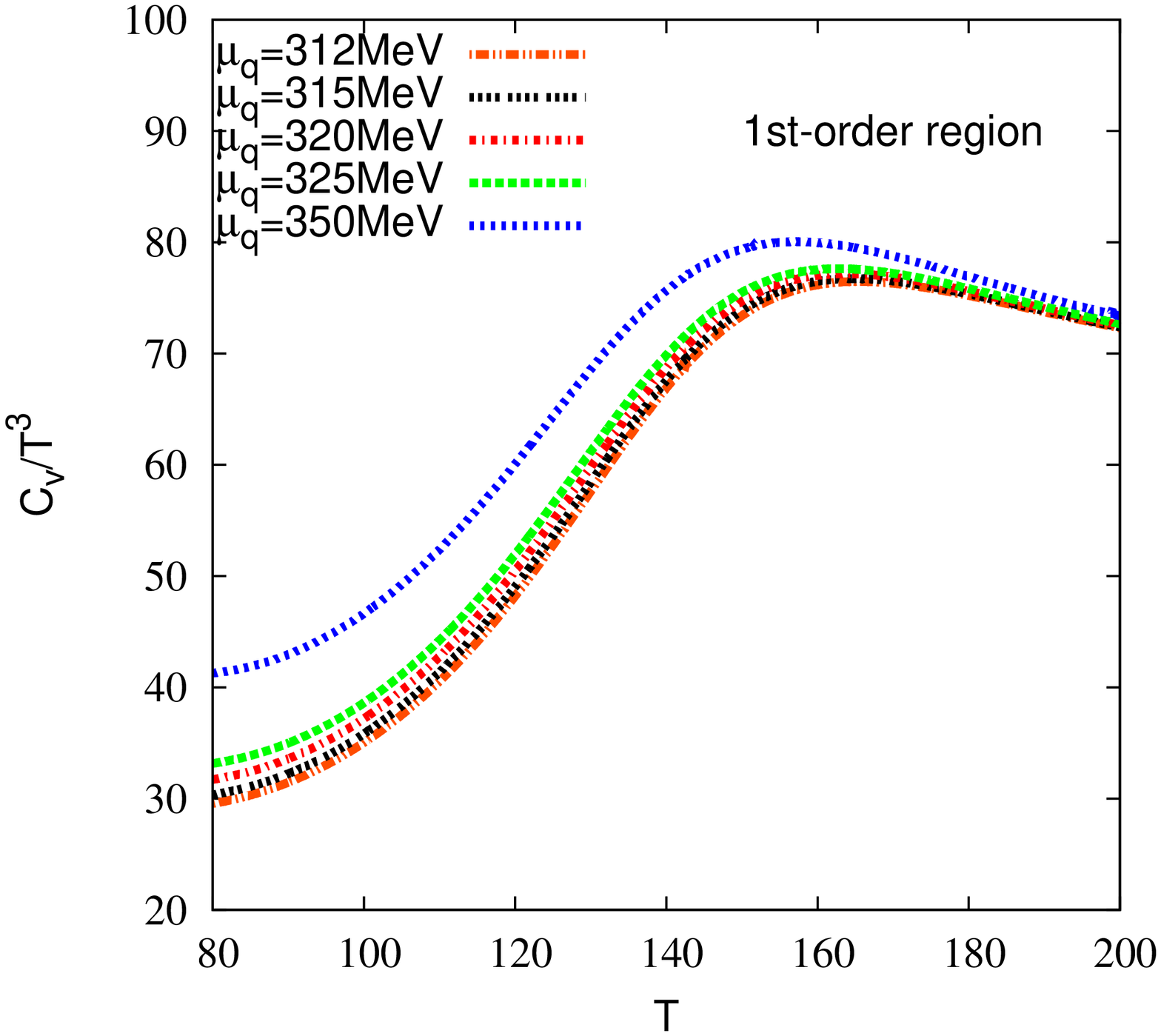}}
 \caption{(color online)$C_V/T^3$ as a function of temperature for different
values of $\mu_q$}. 
\label{Cv}
\end{figure}
In Fig.(\ref{Cv}), we have shown the variation of dimensionless
quantity $\frac{C_V}{T^3}$ with T separately in the regions of
cross-over, CEP, 1st order and beyond, along the $\mu_q$ direction.
As is seen $\frac{C_V}{T^3}$ shoots up around $\mu_q\sim 260 MeV$, but shows
continuous behavior elsewhere. Thus we confirm that the CEP is
expected to exist around $T$=100 - 120 MeV and $\mu_q$=250 - 270 MeV
as inferred from the behavior of $\frac{\eta}{s}$ in the previous
subsection. 

\subsection{Connection with experiments}
In experiments, the observables are studied as a function of center of
mass collision energy ($\sqrt s$). In our thermodynamic studies the
independent variables are $T$, $\mu_B,~\mu_Q$ and $\mu_S$ where
$\mu_B$, $\mu_Q$ and $\mu_S$ are the the baryon, electric charge and
strangeness chemical potentials respectively. So to get the collision
energy dependence one needs to get a parameterization between $\sqrt{s}$
and various thermodynamic variables. Different parameterizations of the
freeze-out conditions as function of $\sqrt s$ are available in the
literature. For a given set of the thermodynamic variables, the
variations in $\sqrt s$ are within 10 $\%$ for different
parameterizations. In the present study we have used the following
parameterization  \cite{Redlich}:
\begin{equation}
 T(\mu_B)~=~a-b\mu_B^2-c\mu_B^4 ~~\&~~
 \mu_{B,Q,S}(\sqrt s)~=~\frac{d}{1+e\sqrt s} 
\end{equation}
where, $a = (0.166 \pm 0.002)GeV,~ b=(0.139\pm0.016)GeV^{-1},
~ c=(0.053\pm0.021)GeV^{-3}$ and $d$ and $e$ are given by :\\
\begin{center}
\begin{table}[!h]
\label{tablesimple}
\begin{tabular}{|c|c|c|}
\hline
&$d[GeV]$&$e[GeV^{-1}]$\\
\hline
B&1.308(28)&0.273(8)\\
\hline
Q&0.0211&0.106\\
\hline
S&0.214&0.161\\
\hline
\hline
\end{tabular}
\end{table}
\end{center}

\begin{figure}[!h]
 {\includegraphics[height=8.5cm,width=6cm,angle=270]{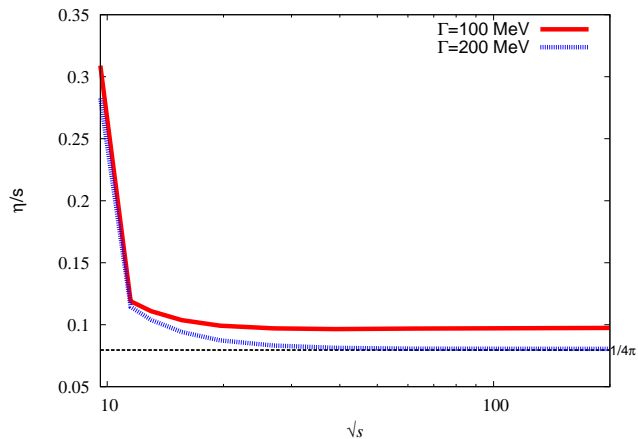}}
 \caption{(color online)$\frac{\eta}{s}$ under different experimental
conditions along the freeze-out diagram}. 
\label{sqrts}
\end{figure}

The variation of $\frac{\eta}{s}$ as a function of $\sqrt s$ along the
freeze-out curve is displayed in Fig.(\ref{sqrts}). We have already seen
that increase in $\Gamma$ results in the decrease in $\eta$. This is
true for non vanishing chemical potential as well. On the other hand, at
very high energies, $\frac{\eta}{s}$ is expected to maintain the
specified lower limit or KSS bound of $\frac{1}{4\pi}$. In
Fig.(\ref{sqrts}), we have plotted specific shear viscosity considering
a spectral width of 100 and 200 MeV. It can be seen that
$\frac{\eta}{s}$ saturates to the KSS bound for $\Gamma$=200 MeV whereas
for $\Gamma$=100 MeV saturation occurs at higher values. So, in the
present study, the stronger interaction (in terms of $\Gamma$) seems to
be necessary to attain the KSS bound.
 
It should be noted that the $\frac{\eta}{s}$ plotted here are for a
system in complete thermodynamic equilibrium at a given value of
temperature and chemical potential. Moreover, along the freeze-out
curve, we always reside in the hadronic phase. So, the value of
$\frac{\eta}{s}$ in Fig.(\ref{sqrts}) corresponds to different
conditions in the hadronic sector. Initially $\frac{\eta}{s}$ decreases
with $\sqrt{s}$ and then saturates at a value close to the KSS bound for
higher $\sqrt{s}$. This is a reflection of the variation of
$\frac{\eta}{s}$ as shown in Fig.(\ref{etabysT-mu0}) and
Fig.(\ref{etabysT}). Our results as shown in Fig.(\ref{sqrts}) is close
to $\frac{\eta}{s}\sim$ 0.09$\pm$0.015 \cite{Lacey,RALacey} as extracted
from the RHIC data for Au+Au collisions. Gavin and Abdel-Aziz \cite{TG}
estimated, from the STAR data analysis, $\frac{\eta}{s}$ to lie in the
range of 0.08-0.3. Another estimate done for top RHIC energies render a
value of $\frac{\eta}{s}$=0.12 \cite{Tribedy} which is also close to our
model results. 

At RHIC, large azimuthal anisotropy of transverse momentum ($p_T$)
spectra, often expressed by the elliptic flow coefficient $v_2$ has been
observed and this is considered to be a signature for the formation of
QGP. The observed transverse momentum spectra of the hadrons and
their centrality dependence
\cite{Prithwish,Hirano,Song,Denicol,Shen,Huovinen} in relativistic heavy
ion collision experiments can be explained satisfactorily using fluid
dynamic descriptions. In fact, the notion of small $\frac{\eta}{s}$ for
QGP arose from analysis of RHIC data through hydrodynamical simulations
\cite{Niemi,Heinz,Romatschke,Shen,Song} which indicate non-viscous fluid
properties for QGP. Though there have been predictions from various
hydrodynamic calculations and simulations regarding 
temperature-independent $\eta$, uncertainties occur because of different
initial conditions used. For example, use of data from Au+Au collisions
at RHIC leads to $(\frac{\eta}{s})_{QGP}~\sim$ 0.08 with MC-Glauber
initializations, whereas that with MC-KLN results in 0.16 \cite{Heinz}.
Moreover, analysis at LHC energies for Pb+Pb collisions provide a bit
higher result $\sim$ 0.2 with various parameterizations incorporating
different initial conditions \cite{Denicol,Tribedy}.

\section{Conclusion}
In the present study PNJL model has been used to calculate shear
viscosity $\eta$ at finite temperature and chemical potential using Kubo
formalism. Effect of spectral width $\Gamma$ has been discussed and
finally constant $\Gamma$ has been used to present rest of our results.
We have also presented our model results for $\frac{\eta}{s}$ for
$T \ne$ 0, $\mu_q$ =0 and $T \ne$ 0, $\mu_q \ne$0 cases. Behavior of 
$\frac{\eta}{s}$ with $T$ and $\mu_q$ has been used to locate the CEP.
This location of CEP has also been validated using the variation of
specific heat $C_V$ with $T$ and $\mu_q$. Finally we have studied the
variation of $\frac{\eta}{s}$ with $\sqrt{s}$ to compare with the
values extracted from the analysis of RHIC and LHC data existing in the
literature.

The salient points of the present study are ::

\begin{itemize}
\item $\eta$ by itself seems to show a gas like behavior and increases
with increasing temperature. It has a strong dependence on the spectral
width $\Gamma$, especially at higher temperatures.

\item At $\mu_q$=0, $\frac{\eta}{s}$, in the present model, seems to
reproduce the pion gas feature at low temperature and QGP features at
high temperatures. Near $T_c$, $\frac{\eta}{s}$ has a minima at the
KSS bound. For $T > T_c$, $\frac{\eta}{s}$ initially increases and then
starts decreasing with increasing temperature showing a liquid like
behavior. Moreover it agrees with the results from interacting pion gas
at low $T$ and lattice data for gluon plasma at high $T$.

\item The effect of non zero $\mu_q$ is found to be interesting. At
$T$=100 MeV, up to $\mu_q$=250 MeV $\eta$ remains almost same as that for
$\mu_q$=0 and then increases sharply to about 0.155 at $\mu_q$=400 MeV
which is more than order of magnitude higher than the $\mu_q$=0 case.
For higher $T$ the value of $\mu_q$, for which $\eta$ retains the
$\mu_q$=0 value, decreases. For example at $T$=200 MeV and 
$\mu_q$=250 MeV, $\eta$ is twice that of its $\mu_q$=0 value.

\item For non-zero $\mu_q$, the minimum value of $\frac{\eta}{s}$ is
found to be higher for larger $\mu_q$ and does not reach the KSS bound
for $\mu_q \ge$ 200 MeV.

\item Variation of $\eta$ across the phase boundary depends strongly on
the nature of phase transition. In the cross over region, 
$\frac{\eta}{s}$ changes continuously with $\mu_q$ whereas it shows a
jump in the first order transition region. The change in the nature
of variation, in going from cross over to first order, may be used to
extract the information of CEP. According to present analysis, CEP
seems to lie in the range $T$=100 - 120 MeV and $\mu_q$= 250 - 270 MeV.

\item $\frac{\eta}{s}$ on the freeze out curve seem to agree with the
values extracted from RHIC and LHC flow analysis. At lower $\sqrt{s}$
hadronic features seem to dominate which may be verified at FAIR
experiments.      
\end{itemize}

It will be interesting to use the present $T$ and $\mu_q$ dependence of
$\frac{\eta}{s}$ for flow analysis to reproduce the data and will be
pursued in our future studies. 

\begin{acknowledgments}
The authors would like to thank CSIR and DST for funding this work.
S.U. thanks Avik Banerjee for few useful discussions regarding the
framework of Kubo formalism. S.U. and K.S. acknowledge Sabyasachi
Ghosh and Sarbani Majumder for useful suggestions.
\end{acknowledgments}


\end{document}